%% file: main.tex
\documentclass[12pt]{iopart}

\usepackage{booktabs}
\usepackage{comment,mfirstuc}
\usepackage{graphicx}
\usepackage{grffile}
\usepackage{algorithm}
\usepackage{algpseudocode}
\usepackage{hyperref}
\usepackage{soul}
\usepackage{listings}
\usepackage{multirow}
\usepackage{dcolumn}
\usepackage{bm}
\usepackage{placeins}
\expandafter\let\csname equation*\endcsname\relax
\expandafter\let\csname endequation*\endcsname\relax
\usepackage{amsmath}
\usepackage{amssymb}
\usepackage{mathtools}
\usepackage[bottom=1.0in]{geometry}
\usepackage{cite}
\usepackage{lineno}
\RequirePackage{xcolor}
\usepackage[misc]{ifsym}
\usepackage{lipsum} 
\usepackage{subcaption}
\usepackage{tcolorbox} 
\usepackage{stfloats} 
\usepackage{siunitx}
\DeclareSIUnit\eVperc{\eV\per\clight}
\DeclareSIUnit\clight{\text{\ensuremath{c}}}

\usepackage{ulem}

\usepackage{xspace}

\usepackage{tcolorbox} 
\usepackage{stfloats} 

\usepackage{colortbl}
\usepackage{array}
\newcolumntype{P}[1]{>{\centering\arraybackslash}p{#1}}
\newcolumntype{M}[1]{>{\centering\arraybackslash}m{#1}}

\hypersetup{ 
    pdfnewwindow=true,      
    colorlinks=true,       
    linkcolor=blue,         
    citecolor=blue,        
    filecolor=blue,      
    urlcolor=blue        
}  

\algblock{Input}{EndInput}
\algnotext{EndInput}
\algblock{Output}{EndOutput}
\algnotext{EndOutput}

\def\be{\begin{eqnarray} &&} 
 
\def\ee{\end{eqnarray}}

\makeatletter
\newcommand{\mainmatter}{%
  \setcounter{footnote}{0}%
  \patchcmd{\@makefntext}{\fnsymbol}{\arabic}{}{}%
  \patchcmd{\@thefnmark}{\fnsymbol}{\arabic}{}{}%
  \def\@makefnmark{\textsuperscript{\arabic{footnote}}}
  \long\def\@makefntext##1{\parindent 1em\noindent
        \hb@xt@1.8em{%
            \hss\@textsuperscript{\normalfont\@thefnmark}}##1}%
}
\makeatother

\newcommand{\addComment}[2]{
  \expandafter\newcommand\csname #1\endcsname[1]{{\bf \color{#2} \capitalisewords{#1}:\,##1}}
  \expandafter\newcommand\csname #1cor\endcsname[2]{{\color{#2} \capitalisewords{#1}:\,\st{##1}{\bf ##2}}}
  \expandafter\newcommand\csname #1color\endcsname{#2}
}
\addComment{cris}{blue} 
\addComment{james}{red}
\addComment{ben}{green}
\addComment{cole}{purple}
\addComment{carlos}{orange}
\addComment{gerald}{brown}

\begin{document}

\providecommand{\keywords}[1]
{
  \small
  \textbf{Keywords:}  {\color{blue}#1
  }
}
\title[\scriptsize{Generalizable Foundation Models for Calorimetry via Mixtures-of-Experts and Parameter Efficient Fine Tuning}]{Generalizable Foundation Models for Calorimetry via Mixtures-of-Experts and Parameter Efficient Fine Tuning}


\author{Carlos Cardona-Giraldo$^{1,*}$, Cristiano Fanelli$^{2,3,4,*}$, James Giroux$^{2,*}$, Cole Granger$^{2,*}$, Benjamin Nachman$^{3,4,*}$, Gerald Sabin$^{1,*}$} 

\address{
$^{1}$ RNET Technologies Inc., Dayton, OH 45429 \\
$^{2}$ William \& Mary, Department of Data Science, Williamsburg, VA 23185, USA\\
$^{3}$ Fundamental Physics Directorate, SLAC National Accelerator Laboratory, Menlo Park, CA 94025,
USA \\
$^{4}$ Department of Particle Physics and Astrophysics, Stanford University, Stanford, CA 94305, USA \\
$^{\star}$ Author to whom any correspondence should be addressed.
}

\ead{{\color{blue}
ccardona@rnet-tech.com,
cfanelli@wm.edu,
jgiroux@wm.edu,
cjgranger@wm.edu,
nachman@stanford.edu,
gsabin@rnet-tech.com
}}

\vspace{10pt}
\begin{indented}
\item[]\today
\end{indented}


\begin{abstract}
Modern particle physics experiments face an increasing demand for high-fidelity detector simulation as luminosities rise and computational requirements approach the limits of available resources. Deep generative models have emerged as promising surrogates for traditional Monte Carlo simulation, with recent advances drawing inspiration from large language models (LLM) and next-token prediction paradigms. In this work, we introduce a generalizable foundation model for calorimetry built on next-token transformer backbones, designed to support modular adaptation across materials, particle species, and detector configurations.
Our approach combines Mixture-of-Experts pre-training with parameter-efficient fine-tuning strategies to enable controlled, additive model expansion without catastrophic forgetting. A pre-trained backbone is trained to generate electromagnetic showers across multiple absorber materials, while new materials are incorporated through the addition and tuning of lightweight expert modules. Extensions to new particle types are achieved via parameter-efficient fine-tuning and modular vocabularies, preserving the integrity of the base model. This design enables efficient, incremental knowledge integration as new simulation datasets become available, a critical requirement in realistic detector-development workflows.
In addition, we demonstrate that next-token calorimeter models are computationally competitive with standard generative approaches under established LLM optimization procedures. These results establish next-token architectures as a viable path toward extensible, physics-aware foundation models for calorimetry and future high-energy physics experiments.
\end{abstract}


%
%
%
%
%

\mainmatter

\input{1_introduction}

\input{2_problem}

\input{3_architecture}

\input{4_analysis}

\input{5_summary}

\section*{Code and Data Availability}

The code used in this work is publicly available at \url{https://github.com/wmdataphys/FM4CAL}. Datasets can be reproduced following the instructions at \url{https://github.com/FLC-QU-hep/getting_high}. Additional information on modification of detector materials is available upon request.



\section*{Acknowledgments}

%
%
%
We thank Vinicius Mikuni for insightful comments and valuable discussions.

This material is based upon work supported by the U.S. Department of Energy, Office of Science, under Small Business Technology Transfer (STTR) Phase I award funded under DOE Contract No. DE-SC0025823. The authors acknowledge William \& Mary Research Computing for providing computational resources and technical support that have contributed to the results reported within this article.

\section*{References}
\bibliographystyle{iopart-num}
\bibliography{biblio}

\clearpage
\input{6_Appendix}

\end{document}

%% file: 1_introduction.tex
\section{Introduction}\label{sec:intro}
%

Modern particle physics experiments are characterized by a growing demand for high-fidelity simulation of detector responses. This trend is expected to intensify as luminosities rise, with estimates projecting that simulation requirements will surpass available computing resources in the near future \cite{adelmann2022new,CERN-LHCC-2022-005,Software:2815292}. 
As a result, new approaches must be explored in which Deep Learning (DL) serves as an efficient and scalable surrogate for Monte Carlo (MC) simulation. In particular, the rapid rise of generative artificial intelligence (AI) in adjacent domains such as Natural Language Processing (NLP) and image generation has generated significant interest within the physics community. Many of these methods can be naturally adapted to model the low-level readout of modern detectors.  Calorimetry is of particular interest, as it often constitutes the primary computational bottleneck in nuclear and particle physics detector simulations. This is driven by the requirement to fully contain high-energy particles, which entails modeling multi-scale processes and extensive cascades of secondary interactions.
In the past, various methods such as Generative Adversarial Networks (GANs) \cite{Paganini_2018,Paganini_2018_accel,Chekalina_2019,Erdmann_2019,carminati2017calorimetry,Belayneh_2020,buhmann2021getting,rehm2021validationdeepconvolutionalgenerative,khattak2021fastsimulationhighgranularity,Bieringer_2022,Rogachev_2023,diefenbacher2023newanglesfastcalorimeter,Giannelli_2024,Simsek_2024}, Variational Auto-encoders (VAEs) \cite{abhishek2022calodvaediscretevariational,cresswell2022calomanfastgenerationcalorimeter,hoque2024caloqvaesimulatinghighenergy,liu2024calovqvectorquantizedtwostagegenerative,smith2024fastmultigeometrycalorimetersimulation}, normalizing flows \cite{Krause_2023,krause2023caloflowiifasteraccurate,Diefenbacher_2023,Buckley_2024,Pang_2024,Erdmann_2025}, and diffusion models \cite{Mikuni_2022,Buhmann_2023_clouds,amram2023denoisingdiffusionmodelsgeometry,diefenbacher2025refiningfastcalorimetersimulations,Mikuni_2024,buhmann2024calocloudsiiultrafastgeometryindependent,Kobylianskii_2024,Favaro_2025,Buss_2026,raikwar2025generalisablegenerativemodelmultidetector,buss2025caloclouds3ultrafastgeometryindependenthighlygranular,favaro2025fastaccurateprecisedetector,buss2026allshowersmodelcalorimetershowers} have been used to create these efficient surrogate models for calorimeter simulations. More exhaustive collections of these frameworks can be found in \cite{hepmllivingreview}, with benchmarking of various models found in \cite{CaloChallenge,buss2025physicsbenchmarkhighlygranular}. Following the success of Large Language Models (LLMs), additional methods of calorimeter simulation have been proposed utilizing a next-token methodology \cite{birk2025omnijet} where showers are generated sequentially hit by hit. Next-token models have shown both exceptional generation ability, and scalability in terms of both compute and ability to learn from data. Specifically, transformer models are generally well known for their ability to scale with pre-training dataset size. 
Moreover, transformer-based models exhibit strong fine-tuning efficiency, as demonstrated by the widespread deployment of few-shot learning techniques in NLP.

In foundation models (FM), fine-tuning efficiency—measured by the data and time required for adaptation—can substantially reduce computational cost in particle physics applications from both Central Processing Unit (CPU) and Graphics Processing Unit (GPU) perspectives. Detector optimization workflows \cite{diefenthaler2024ai,cisbani2020ai}, for example, are typically dominated by large-scale \textsc{Geant4} \cite{AGOSTINELLI2003} simulations of particle interactions in candidate materials. 
In this work, we consider absorber materials for calorimetry such as tungsten (W) and tantalum (Ta), which require extensive simulations to generate datasets for downstream reconstruction and performance studies. If a new material, such as lead (Pb), is introduced during optimization, the conventional approach requires another large, CPU-bound simulation campaign.
A foundation model paradigm like the one we propose alleviates this burden. Simulation samples generated for the initial materials can be used both to train the model and to support the optimization pipeline. When a new material is introduced, only a modest number of additional simulations may be needed to fine-tune the pre-trained model. The adapted model can then generate high-fidelity detector-response samples efficiently on GPUs, amortizing the original CPU cost and yielding orders-of-magnitude reduction in required simulation.
From a GPU standpoint, efficient fine-tuning reduces the training data and iterations required for adaptation, shortening wall-clock time and lowering energy consumption. This shift—from repeated CPU-intensive simulation to targeted fine-tuning and GPU-based generation—enables a scalable and computationally sustainable detector optimization strategy.

Fine-tuning alone does not provide a complete solution though; its downstream implications must be evaluated in light of the chosen adaptation strategy. While full fine-tuning is arguably the most expressive approach, it introduces the risk of misalignment or catastrophic forgetting within the base model \cite{kirkpatrick2017overcoming}. For instance, a model trained to generate photons in a given detector configuration may, after full fine-tuning to electrons in the same setup, lose its ability to faithfully reproduce the original photon distribution. In extreme cases, this degradation may be effectively irrecoverable. In certain cases, where recovery of the base model is not a concern, full fine-tuning is an invaluable strategy. For example, in \cite{gaede2025crossgeometrytransferlearningfast} the authors show that information can be transferred across detector geometries, ultimately increasing training efficiency and/or model performance.

Ideally, fine-tuning (or more generally adaptation) should be additive or \textit{modular} in nature. Such a design ensures that newly introduced capabilities can be selectively enabled or disabled, preserving the integrity of the original model parameters. In this way, extensions to new particle types or detector configurations do not perturb the base model, but instead augment it in a controlled and reversible manner.
Ultimately, we seek a model that, following pre-training, can dynamically extend its capabilities to new configurations as additional datasets become available without risking misalignment with the original base model. 
%
%
A related approach was recently proposed in \cite{nguyen2026differentiablesurrogatedetectorsimulation}, where Denoising Diffusion Probabilistic Models were used to construct a differentiable surrogate for electromagnetic calorimeter simulations conditioned on detector-design parameters. 

We envision a larger program of detector design, calibration, operation, and analysis for a diverse set of instruments. The ultimate goal is a foundation model that would facilitate rapid adaptation for maximum output in both fundamental and applied science, medicine, and technology.
The central objective of this work is to design a generalizable foundation model for calorimetry, built on next-token prediction backbones, that is \textit{transferable} across materials and detector design configurations. The main contributions of this work are as follows:

(i) We construct a pre-trained backbone model capable of generating photons in multiple materials under one model through a \textit{Mixture-of-Experts} (MoE), adapting existing FMs \cite{giroux2025towards} from other areas of physics.

(ii) We show efficient and modular fine-tuning to additional materials through the addition and tuning of only a single expert.

(iii) We show efficient and modular fine-tuning to additional particles through \textit{Parameter Efficient Fine-Tuning} methods and modular vocabularies.

(iv) We show that next-token prediction models are computationally competitive with standard generative approaches using established LLM optimization techniques.

In practical development settings, comprehensive simulation data is seldom available at the time of pre-training. Accordingly, the model must support incremental integration of new knowledge, enabling a controlled and principled expansion of its prior, while preserving previously learned behaviors. This philosophy underpins the structure of the sections that follow. In Sec.~\ref{sec:applications}, we describe the detector configuration and the corresponding datasets used throughout this study. Sec.~\ref{sec:architecture} details the model architecture, emphasizing its key components and their alignment with the proposed design principles. In Sec.~\ref{sec:results}, we present experimental evaluations of both pre-trained models and subsequent fine-tuning strategies. Finally, Sec.~\ref{sec:summary} concludes with a summary of our findings and broader implications.

%% file: 2_problem.tex
\section{Datasets}\label{sec:applications}


The datasets used in this work replicate those introduced in~\cite{birk2025omnijet,buhmann2021getting}, which simulate electromagnetic calorimeter showers for the International Large Detector (ILD)~\cite{ild2020international}. The ILD is a proposed detector concept for the International Linear Collider (ILC)~\cite{bambade2019international}, an electron-positron collider designed to operate at a center-of-mass energy of $\SI[per-mode=symbol]{250}{\giga\eV}$. Its electromagnetic calorimeter, the Si-W ECAL~\cite{suehara2018performance}, comprises 20 tungsten (W) absorber layers of $\SI{2.1}{\milli\meter}$ thickness, followed by 10 additional tungsten layers of $\SI{4.2}{\milli\meter}$ thickness. Absorber layers are interspersed with $\SI{0.5}{\milli\meter}$ silicon (Si) sensor layers, segmented into $\SI{25}{\milli\meter}^2$ cells. In addition to tungsten, we construct alternative datasets in which the absorber material is replaced with tantalum (Ta) and lead (Pb). These three materials span a range of radiation and interaction properties relevant to realistic detector design trade-offs, making them representative choices for studying material-dependent shower modeling.

Following~\cite{birk2025omnijet,buhmann2021getting}, we uniformly simulate electromagnetic showers with incident energies in the range $10$--$\SI{100}{\giga\eV}$. The detector response is projected onto a voxelized three-dimensional grid of size $30\times30\times30$. This procedure is performed for both photons and electrons, and for each absorber material considered. Consistent with the discrete tokenization scheme of our method, we operate directly on the voxelized representation and store sparse encodings of voxels with non-zero deposited energy. Each dataset contains approximately 950k samples, partitioned into 760k for training, 95k for validation, and 95k for testing, mirroring the data split of~\cite{birk2025omnijet}.

We emphasize that the dataset size and composition are intentionally kept consistent with prior work \cite{birk2025omnijet} to enable direct and fair comparison across methods when permitted. As a result, the scale of pre-training data used in this study is not reflective of the full capacity or intended deployment regime of the proposed models. In practice, we expect performance to improve with substantially larger and more diverse simulated datasets, and future work will explore this scaling behavior in detail, particularly in the context of foundation model pre-training and downstream adaptation.


%% file: 3_architecture.tex
\section{Architecture}\label{sec:architecture} 

\begin{figure}[ht]
    \centering
    \includegraphics[trim={1mm 0mm 42mm 0mm},clip,width=\textwidth]{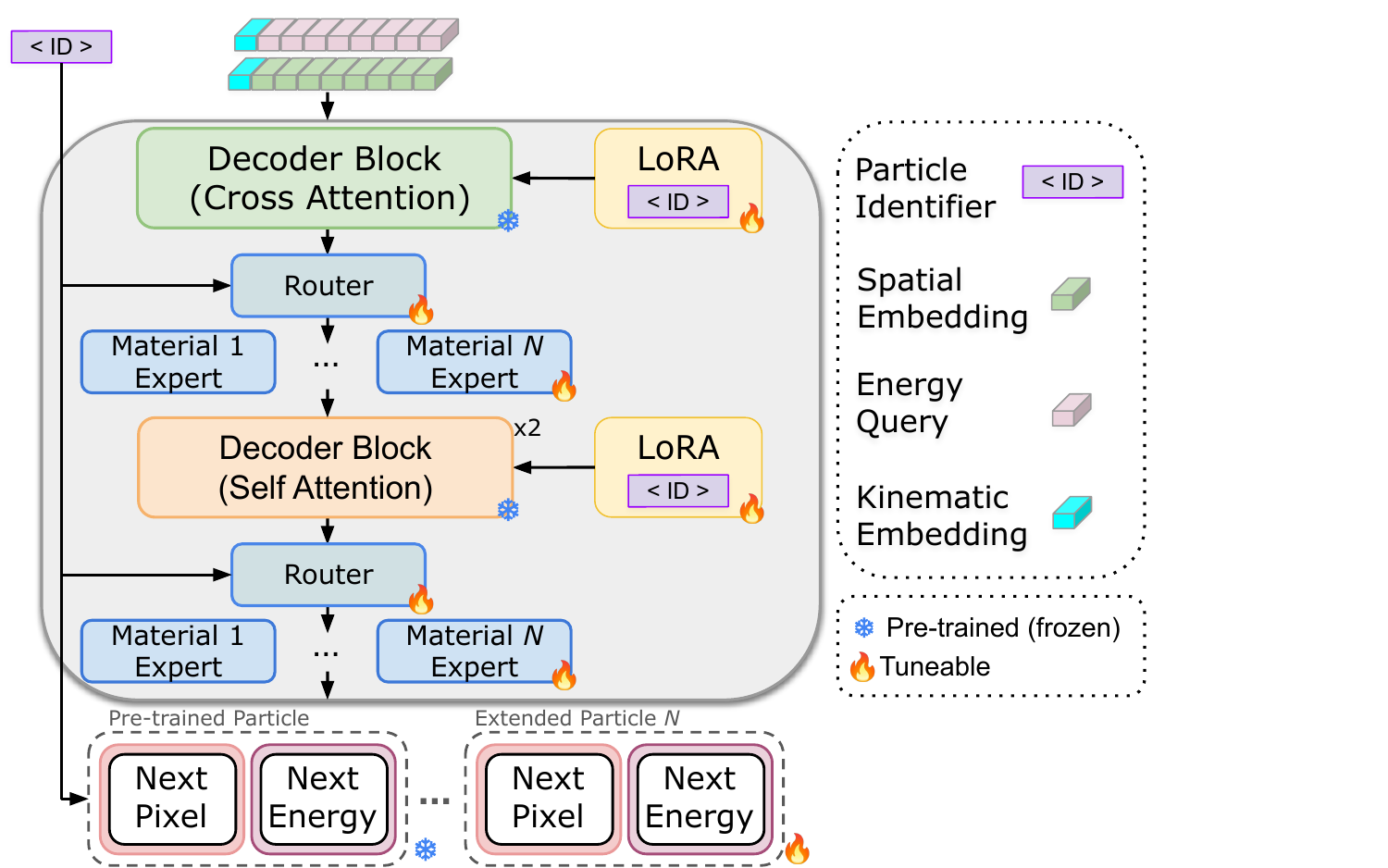}
    \caption{\textbf{Overview of the modular architecture for material and particle-species adaptation}: 
The framework utilizes a core transformer backbone consisting of cross-attention and self-attention decoder blocks that remain frozen during secondary adaptation phases. Material extensibility is achieved through a Mixture-of-Experts (MoE) layer where a router directs inputs to specialized modules, allowing for the addition of new materials by fine-tuning only a singular new expert. When transitioning to different particle species, the model employs a parameter-efficient strategy using LoRA modules and expanded particle-specific vocabulary heads for pixel and energy prediction while the base photon model parameters remain static. For subsequent material expansion of an adapted model, a new expert is integrated while the previously tuned LoRA and vocabulary components are frozen to preserve the learned particle-specific features. The system is conditioned throughout these stages by a combination of spatial, kinematic, and energy query embeddings alongside a unique particle identifier.}
\label{fig:model}
\end{figure}

In what follows, we describe the architecture in a step-by-step manner. The presentation follows the natural sequence of the model construction: we begin with the definition of the base model and its core components, and then introduce the additional modules used during the adaptive phases. A schematic overview of the full architecture is shown in Fig.~\ref{fig:model}.

\subsection*{Tokenization of Space and Energy}

As discussed in \cite{giroux2025towards}, the pixelated nature of readout systems does not require an additional tokenization strategy along the spatial dimension; each discrete pixel index corresponds uniquely to a fixed $(x, y, z)$ readout coordinate, producing a spatial vocabulary of size 27k. 
Along the energy dimension, which is inherently a continuous variate, we apply the same linear binning technique developed in \cite{giroux2025towards} in which we discretize the energy at some fraction of the readout resolution. Given the developmental nature of the ILD detector \cite{ild2020international} no energy resolution is used during simulation. Therefore, we choose a fractional energy bin such that our two vocabularies are approximately equal in size for convenience's sake. The resulting energy vocabulary consists of 25k unique tokens that cover the interval $10^{-15}\,\SI[per-mode=symbol]{}{\mega\eV}$ to $\SI[per-mode=symbol]{35}{\mega\eV}$. This removes the need for optimization of loss weighting parameters. 

\subsection*{Energy Queries Space}


In our approach, energy acts as the guiding principle for sequence generation. Specifically, sequences are ordered by energy in descending magnitude and guided forward in both deposition amount and spatial location by the initial particle energy. This choice naturally suggests a self-consistent generation strategy in which periodic checks—optimized for different particle types and sequence lengths—ensure that the reconstructed energy of the generated sequence matches the expected value. We leave the development of such strategies for future work. In the present work, the current generation is purely open-loop with respect to total energy conservation.

We first construct energy and spatial embeddings via independent learnable projections from their respective vocabularies,
$ \phi_E : \mathcal{V}_E \rightarrow \mathbb{R}^d, \phi_S : \mathcal{V}_S \rightarrow \mathbb{R}^d,$ where $d = 256$ denotes the embedding dimension.
%
In addition, the initial particle energy is embedded through a linear projection $\phi_I : \mathbb{R} \rightarrow \mathbb{R}^d$,
and prepended as conditioning context to both the energy and spatial embedding streams, Eq. \ref{eq:tokens}.

\begin{align}\label{eq:tokens}
    \text{spatial} &\rightarrow \{I,\text{SOS}_p,p_1, \ldots , p_{n}, \text{EOS}_p\} \notag \\
    \text{energy} &\rightarrow \{I,\text{SOS}_e,e_1, \ldots , e_{n}, \text{EOS}_e\}
\end{align}

Our base architecture directly inherits the dual sequence transformer framework introduced in prior work \cite{giroux2025towards}. In particular, information is fused through a cross multi-head cross-attention (CMHCA) block, where Query ($Q$) projections are obtained from the energy embeddings and Key ($K$) and Value ($V$) projections from the spatial embeddings, yielding a spatial-energy aware latent representation. This representation is subsequently processed by standard Multi-Head Self-Attention (MHSA) \cite{vaswani2017attention} transformer blocks. We remove the $\ell_2$ normalization of the $Q$ and $K$ matrices \cite{henry2020query}, as it was found to not improve generalization in this case.

To alleviate the constraints imposed by fixed sequence lengths, we move away from learned positional embeddings (LPE) and instead adopt Rotary Positional Embeddings (RoPE) \cite{su2023roformerenhancedtransformerrotary}. RoPE encodes positional information by applying a position-dependent rotation to the query and key representations, introducing relative positional structure directly within the attention mechanism. In contrast to LPE, which injects positional information additively into token embeddings, RoPE acts multiplicatively on the projected queries and keys. Specifically, the attention score between tokens at positions $m$ and $n$ is computed as

\begin{equation}
\label{eq:RoPE}
Q_m^{\top} R_{n-m} K_n 
\end{equation}

where $R_{n-m}$ is an orthogonal rotation matrix constructed from position-dependent angles $(n-m)\theta^{-2i/d}$ defining geometrically spaced frequencies across the embedding dimension, with $i$ being the dimension index. Standard LLMs typically employ a $\theta$ base of at least 10,000 \cite{su2023roformerenhancedtransformerrotary}; our study found a significantly lower value of 1,000 to be optimal, this being primarily driven by significantly shorter sequence lengths observed in our data.
The RoPE formulation naturally incorporates relative positional dependencies while preserving compatibility with variable sequence lengths. The latter is a crucial characteristic of our model given that different particle species or material densities can cause increased or decreased interaction (number of hits) within the detector. We specifically use Partial RoPE \cite{barbero2025round,ji2025economicalinferenceenablingdeepseeks,deepseekai2024deepseekv2strongeconomicalefficient,black2022gptneox20bopensourceautoregressivelanguage} in which we only rotate the first 50\% of embedding dimensions. Partial RoPE is similar in nature to p-RoPE \cite{barbero2025round} in which the authors find that partially removing rotation better balances semantic and positional information within the attention. In addition, we incorporate a per-token geometric bias that encodes the depth of the shower within the calorimeter. This is implemented as a normalized longitudinal coordinate, \textit{e.g.} $p_i / N_z$, where $N_z = 30$ denotes the total number of discrete $z$-layers.
While our sequences are not long-context in the traditional LLM sense, they are long in relation to the value of $\theta$ we found suitable. As such, we encounter similar issues as those studied in \cite{wang2024precisionmeetspositionbfloat16} in which 16-bit precision causes excess error accumulation as sequence lengths increase. This produces ``artifacts'' in generations, commonly manifested as over represented values of sequence lengths (number of hits). Therefore, we deploy our model at full 32-bit precision and leave more efficient utilization of mixed precision for future studies, or use of AnchorAttention \cite{wang2024precisionmeetspositionbfloat16}.

\subsection*{Material Modeling with Fixed-Routing Mixture-of-Experts}

As shown in \cite{giroux2025towards}, a discrete valued conditional state (\textit{e.g.}, particle type) can be encoded through a Mixture-of-Experts (MoE), in which fixed, class-wise routing allows disentanglement of specific conditional features, while allowing weight sharing across the majority of the model. In our case, we extend this idea to particle, material combinations such that the each expert maps to a unique combinatorial state through a router, Eq. \ref{eq:MoE}, where $N_E = \text{\#(experts per class}) \times \text{\#(materials)} \times \text{\#(particles)}$. In this study, we employ a single expert per class. If multiple experts were to be assigned to a given class, therefore requiring a learned routing, an auxiliary loss term could be introduced to encourage balanced utilization among the experts within that class~\cite{giroux2025towards}.

\begin{equation}\label{eq:MoE}
    z = \sum_{i=1}^{N_{E}} R_i(x) E_i(x), \; \; R_i(x) \in \mathbb{R}^{s \times 1}, \; \; E_i(x) \in \mathbb{R}^{s \times d}
\end{equation}

Naively, one may assume adding additional experts increases overall parameter count at the cost of utility. However, given that routing is fixed, the active parameter count at inference also remains fixed and therefore has no direct limitations in terms of scalability. Explicitly, the inference compute cost remains constant even as the model extends its prior knowledge through fine-tuning. Furthermore, we show that our expert strategy exhibits excellent fine tunability in which we can achieve reasonable approximations of new materials in limited samples. 

\subsection*{Particle Adaptation via Low-Rank Updates and Modular Projections} 

In specific cases, \textit{i.e.}, when introducing new materials for particle species already observed during pretraining—adaptation requires only the addition and tuning of a single expert. Conceptually, this corresponds to a modulation of token probabilities induced by distributional shifts associated with differing radiation lengths. In such scenarios, the underlying token interactions remain structurally consistent, and the expert primarily adjusts marginal likelihoods in the token space.

In contrast, this strategy does not fully generalize to new particle species by itself, where the interaction structure between tokens can change fundamentally. Consider electrons as an illustrative example. Although electromagnetic shower development for electrons shares similarities with photons (\textit{i.e.}, photons undergo pair production), electrons exhibit pronounced differences in transverse depth profiles. In the context of a next-token prediction framework, this manifests as a fundamental shift in the relationship between the \textit{SOS} token and the initial energy deposition pattern. To account for these structural changes, additional modular capacity is required. We introduce this through Low-Rank Adaptations (LoRA) \cite{hu2022lora} applied within the attention blocks. Specifically, low-rank updates are applied to the $Q$, $K$, and $V$ projections, as well as the output projection. LoRA provides a computationally efficient additive modulation of pretrained weights via low-rank factorization (Eq.~\ref{eq:LoRA}), where $B \in \mathbb{R}^{d \times r}$ and $A \in \mathbb{R}^{r \times k}$ with $r \ll \min(d,k)$. In practice, we find that a rank of $r=128$ provides sufficient capacity to modulate attention relationships when adapting to new particle species. 

\begin{equation}\label{eq:LoRA}
    h = W_0x + \Delta Wx = W_0x + BAx
\end{equation}

For new particle species, the problem is better framed as domain adaptation rather than conventional fine-tuning. While LoRA provides the capacity to modulate attention relationships—thereby reshaping conditional dependencies and sequence generation dynamics—it operates as a constrained, low-rank perturbation of the pretrained weights. As such, it can reconfigure how existing representational structure is used, but it does not independently redefine token-space probabilities at the embedding level.

Finally, we must account for fundamental shifts in the token probability space induced by particle-dependent vocabularies. A naive extension would apply low-rank adaptation directly to the output projection, analogous to Eq.~\ref{eq:LoRA}. However, effective modulation of large vocabulary matrices typically requires substantially higher ranks, which increases inference cost while diminishing the efficiency benefits of low-rank adaptation.\footnote{Empirically, we observe that $r=128$ yields performance comparable to independent vocabulary heads.}

Instead, we introduce particle-wise output heads selected through a conditional state. This approach allows each particle species to parameterize its own probability space directly, rather than approximating large vocabulary shifts through high-rank corrections to a shared projection. In contrast to low-rank updates on the output matrix, these heads provide explicit full-rank flexibility where it is most required, while remaining computationally efficient at inference. Moreover, we only need to train these vocabularies once per particle as they can be used in conjunction with their frozen LoRA, and fine-tuned through only the addition of an expert.

Taken together, these components form a fully additive and modular adaptation strategy. Experts address distributional shifts, LoRA captures structural adjustments in attention dynamics, embedding modulation layers conservatively adapt input representations, and conditional output heads model particle-specific vocabularies. Crucially, because adaptation is implemented through strictly additive modules applied on top of a frozen backbone, the base model parameters are never modified; catastrophic forgetting of previously learned behaviors is therefore prevented by construction.

%% file: 4_analysis.tex
\section{Material and Particle Transferability: Analysis and Results}\label{sec:results}

In this section, we validate our methodology in a step-wise fashion consistent with the order in which components were introduced:
%
%
\noindent (i) First, we compare our pre-trained model—capable of generating photons in both tungsten and tantalum via expert routing—with existing autoregressive approaches on the same dataset. We also demonstrate the validity of our conditional generation scheme with respect to particle kinematics. Notably, our approach can be readily extended to these existing methods.
\noindent (ii) Second, we will show the capability of the model to be extended to additional materials, \textit{i.e.}, photons in lead, via the addition of a singular expert. 
%
\noindent (iii) Third, we demonstrate that our model can be transferred not only across detector materials but also across particle species through PEFT methods and modular vocabularies. For consistency and to enable a direct comparison, our analysis strictly follows the methodology of~\cite{birk2025omnijet} in terms of metrics. 

\subsection*{Comparison with other Autoregressive Models}

\begin{figure}[b!]  
    \centering
    \includegraphics[trim={32mm 15mm 40mm 28mm},clip,width=\textwidth]{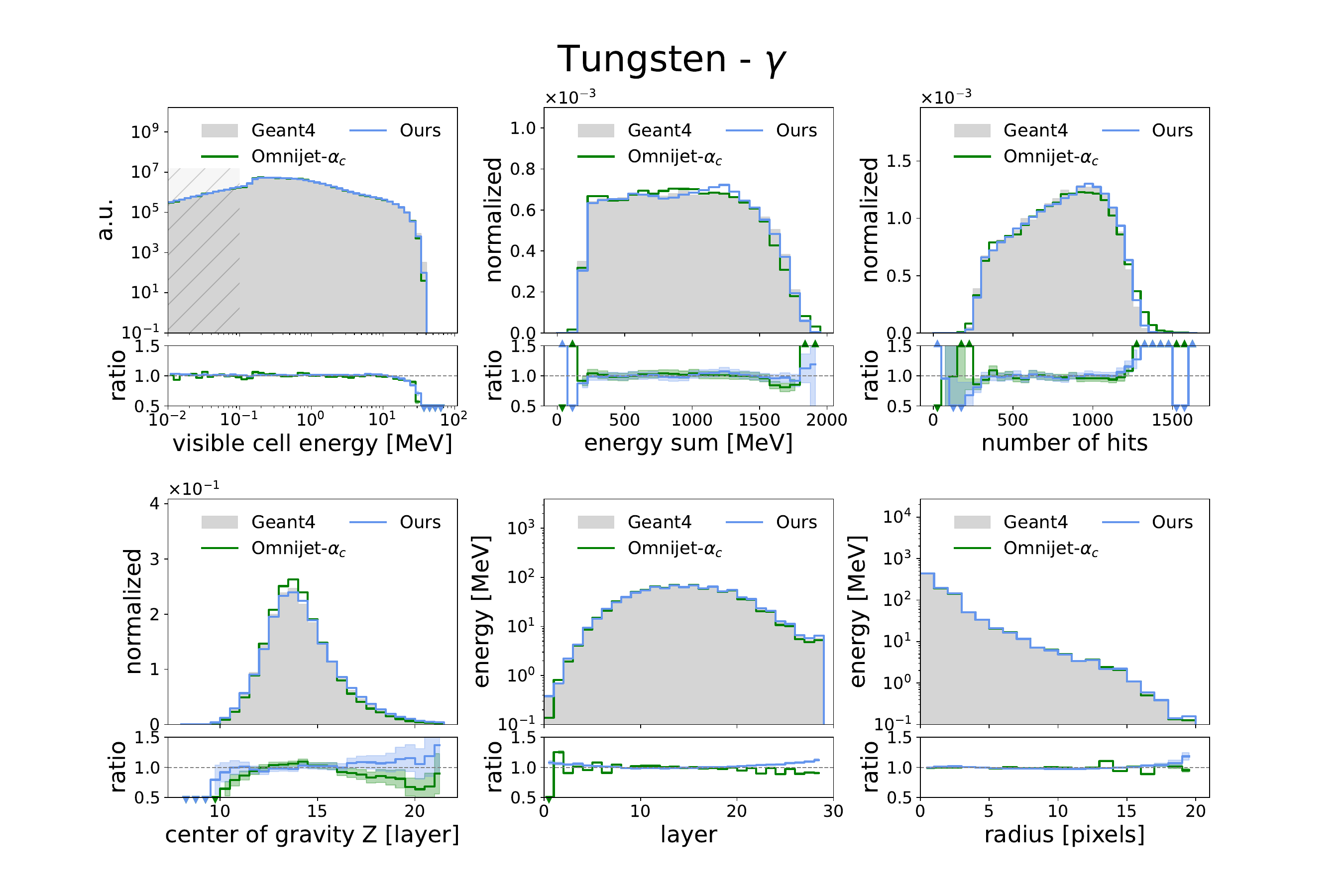}
    \caption{\textbf{Validation of generative shower modeling for photons in tungsten:} 
    Distribution level comparison between ground truth \textsc{Geant4} reference sample (gray shaded), against the proposed method (blue) and the Omnijet-$\alpha_c$ baseline (green) for photons in tungsten.
    The top row depicts observables including visible cell energy (a.u.), total energy sum, and total number of hits. 
    The bottom row depicts spatial shower profiles including the longitudinal center of gravity ($Z$), energy deposition per layer, and radial energy distribution. 
    The lower panels show the ratio of each model to the \textsc{Geant4} reference, where a ratio of $1.0$ (dashed line) indicates perfect agreement. A $3\sigma$ uncertainty profile is provided in the ratio plots, accounting for statistical error in both the generated, and ground truth sample.}
    \label{fig:gamma_W}
\end{figure}

\begin{figure}[!]
    \centering
    \includegraphics[trim={32mm 15mm 40mm 28mm},clip,width=\textwidth]{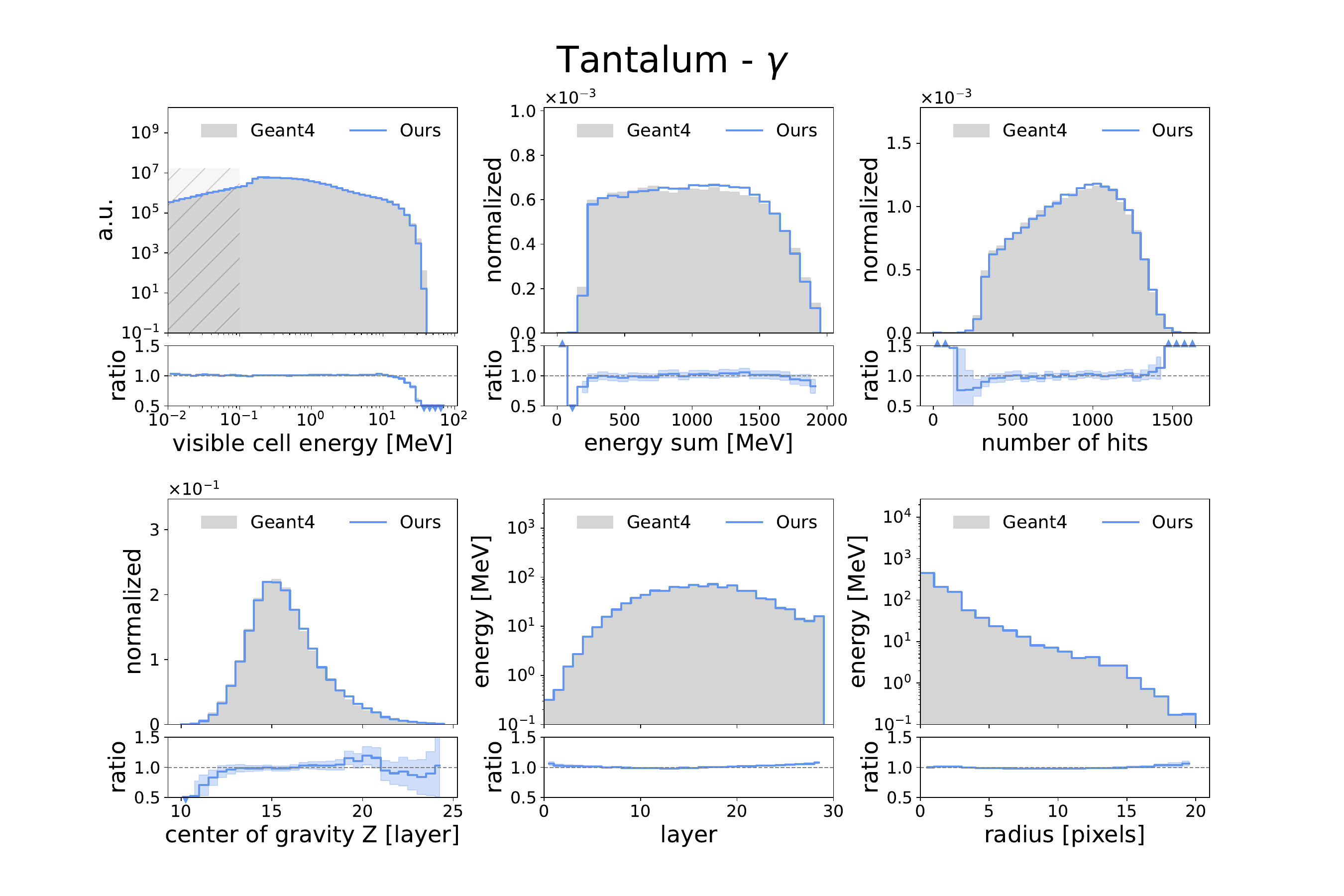}
    \caption{\textbf{Validation of generative shower modeling for photons in tantalum:} 
    Distribution level comparison between ground truth \textsc{Geant4} reference sample (gray shaded) and the proposed method (blue) for photons in tantalum.
    }
    \label{fig:gamma_Ta}
\end{figure}

Figures \ref{fig:gamma_W} and \ref{fig:gamma_Ta} provide a distribution-level evaluation of photon showers in tungsten and tantalum, respectively. We compare our proposed method (blue) against the \textsc{Geant4} ground truth (gray shaded) across the full kinematic range. In the tungsten evaluation (Figure \ref{fig:gamma_W}), we include Omnijet-$\alpha_c$ (green) as a baseline for autoregressive models.

Notably, the tantalum results (Figure \ref{fig:gamma_Ta}) showcase only our model, as Omnijet-$\alpha_c$ is limited to the single material (tungsten) used during its training. Our architecture, however, successfully learns both materials simultaneously during the pre-training phase. This comparison demonstrates that our model can scale to include additional materials without a loss of fidelity, maintaining high-resolution agreement with the ground truth where baseline models require separate, material-specific instances.

The top row (left to right) illustrates  visible cell energy, total energy sum per shower, and hit multiplicity. The bottom row characterizes the spatial shower evolution, depicting the longitudinal center of gravity ($Z$), the average energy deposition per calorimeter layer, and the radial energy profile relative to the shower centroid.


We note good agreement from our method with respect to the reference \textsc{Geant4} population, and Omnijet-$\alpha_c$,  with the primary regions of disagreement coming at the tails of distributions. The reduced sample density in these regions inherently limits the model's ability to learn robust probability distributions for these token sets. Furthermore, this suggests that without extensive pre-training, the transformer backbone's full representational capacity remains underutilized. 
%
%
Within our expert-based conditional generation strategy, the model demonstrates a strong ability to capture distributional shifts, \textit{e.g.}, transverse depth modulation arising from differences in radiation length. It also correctly captures variations in shower multiplicity across materials, with tantalum producing a slightly broader distribution.

\begin{figure}
    \centering
    \includegraphics[trim={32mm 15mm 40mm 28mm},clip,width=\textwidth]{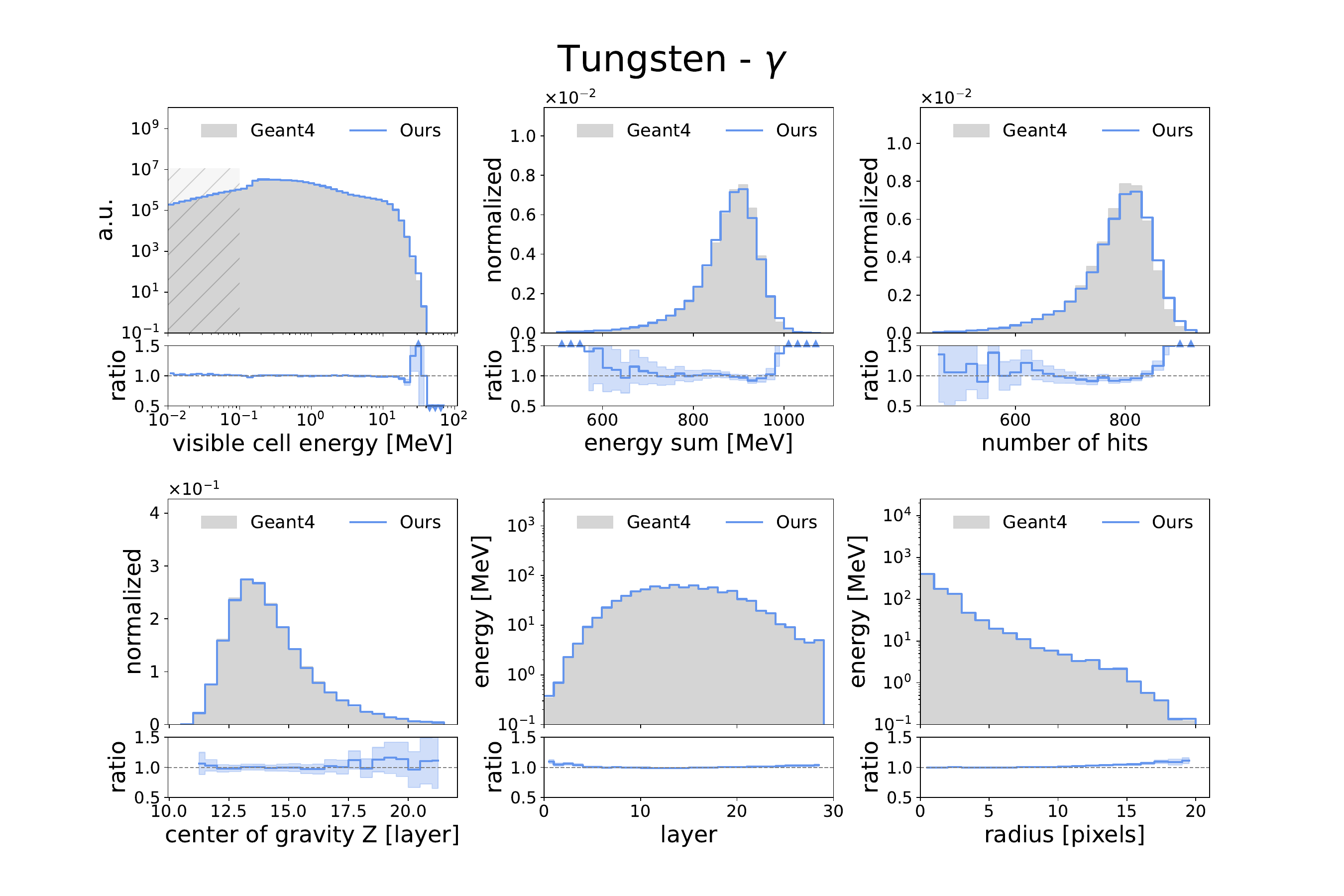}
    \caption{\textbf{Validation of energy kinematic conditioning through prepended context at 50 GeV}:
    Distribution level comparison between ground truth \textsc{Geant4} reference sample (gray shaded) and the proposed method (blue) at a fixed initial energy of $\SI[per-mode=symbol]{50}{\giga\eV}$ for photons in tungsten. 
    }
    \label{fig:gamma_W_50GeV}
\end{figure}

Given that our model utilizes a prepended context conditioning scheme for initial energy, further evaluation is provided in Fig. \ref{fig:gamma_W_50GeV}, in which we show distribution-level evaluations at a fixed value of $\SI[per-mode=symbol]{50}{\giga\eV}$ (\textit{e.g.}, the center of the kinematic range) for $\sim$ 60k photons in tungsten. We remind the reader that our model is trained continuously over the entire kinematic range ($10-\SI[per-mode=symbol]{100}{\giga\eV}$), and therefore evaluations at fixed initial energies serve as stringent assessments of the model’s interpolative stability. Such tests probe the model’s representational fidelity in regions where the training data density is inherently sparse relative to the continuous phase space. The plots follow an identical structure as those prior, with the top row visualizing visible cell energy, energy sum per shower, and multiplicity (left to right). The bottom row visualizes the center of gravity of showers along the z-axis, the average shower energy per layer, and the average radial distance from shower centroid. Additional plots of generations at $\SI[per-mode=symbol]{20}{\giga\eV}$ and $\SI[per-mode=symbol]{80}{\giga\eV}$ can be found in \ref{app:fixed_point}.

From inspection of Figure \ref{fig:gamma_W_50GeV}, we note agreement within uncertainty across all distributions, with increasing disagreement towards lower density regions. We expect that increasing the size of the pre-training dataset will further improve the capability of our prepended strategy, as demonstrated in \cite{giroux2025towards} where more complex, and higher dimensional kinematic correlations are accurately captured.


\subsection*{Extension to Different Materials}

In what follows, we demonstrate that the proposed architecture is both readily extensible and efficiently adaptable to new materials (\textit{e.g.}, photons in lead). Extension is achieved by introducing a single additional expert. All transformer blocks and vocabulary projection layers remain frozen. The parameters of the newly introduced expert are initialized from the previously trained tantalum expert. Expert initialization may be user-defined (\textit{e.g.}, guided by physical priors) or randomly assigned. In our studies, both strategies ultimately converge to comparable solutions; however, informed initialization consistently leads to faster convergence during fine-tuning.

\begin{figure}[!]
    \centering
    \includegraphics[trim={32mm 15mm 40mm 26mm},clip, width=\textwidth]{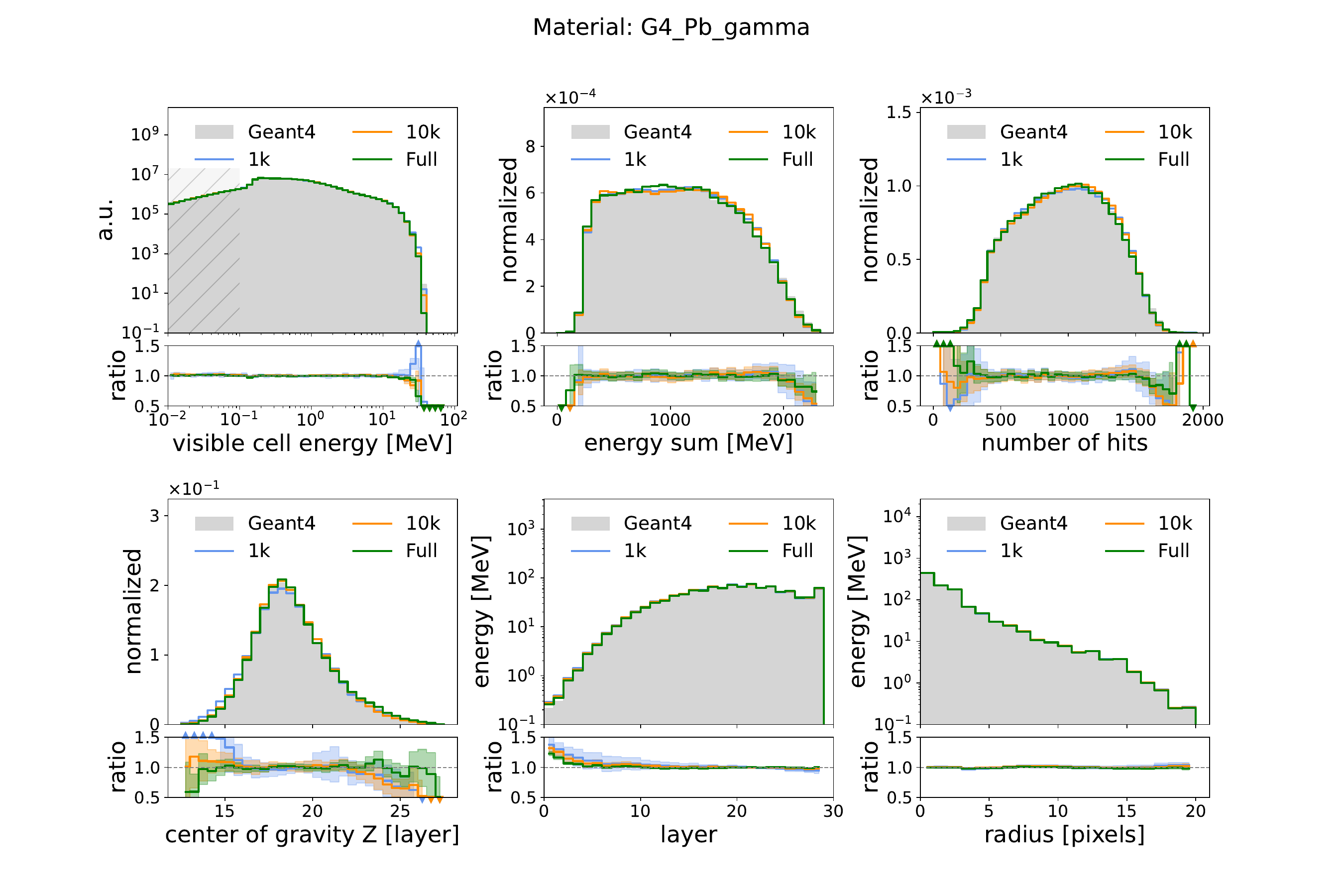}
    \caption{\textbf{Fine-tuning efficiency for photon showers in lead across varying sample sizes:}
    Comparison between the ground truth \textsc{Geant4} reference (black) and the fine-tuned model at $1\text{k}$ (blue), $10\text{k}$ (green), and full (red dashed) sample sizes.
    Lower panels display the ratio to the \textsc{Geant4} reference, where a ratio of $1.0$ (dashed line) indicates perfect agreement. Shaded areas in the ratio plots depict the statistical uncertainty from data samples in quadrature with the uncertainty from bootstrapped fine-tunings, represented at the $3\sigma$ level. Results demonstrate that freezing the base tantalum/tungsten MoE backbone and updating only the lead-specific expert allows for high-fidelity adaptation even with limited training data.}
    \label{fig:gamma_Pb}
\end{figure}

To assess adaptation efficiency, we conduct $k=5$ bootstrapped fine-tuning experiments using subsets of 1k and 10k photon samples in lead. Each bootstrap iteration is initialized with a distinct random seed, yielding independent and unique data selections that span the full phase space. Following fine-tuning, each resulting model is evaluated on the complete test set, from which we compute the mean bin contents and their corresponding standard deviations.
The standard deviation across iterations is then propagated to the ratios and combined in quadrature with the statistical uncertainties of both the generated and ground-truth samples. To save computation, we make an assumption that at our full dataset size, all models converge to approximately identical solutions and therefore the error is reduced to only the statistical uncertainty on the testing samples. The resulting distributions are shown in Figure~\ref{fig:gamma_Pb}, where we present the same distribution-level observables introduced previously. 

From inspection of Figure \ref{fig:gamma_Pb}, we are able to draw two promising conclusions. First, even with relatively low samples seen during fine-tuning (\textit{e.g.}, 1k samples), our model is able to produce generations consistent (within uncertainty) with models that see the entirety of the lead data sample. Such a result is potentially crucial in specific applications where simulating large populations is not feasible, or users have limited amounts of GPU time for training models. 
%
Second, we extend our model to an additional material without introducing misalignment or catastrophic forgetting in the base model. In contrast to full fine-tuning—where the pretrained weights for existing materials (tungsten and tantalum) may shift toward distributions associated with the new material—our approach preserves the original representations. This strategy prevents degradation of the base model while enabling adaptation to additional materials.

\subsection*{Extension to Different Materials and Particles}

\begin{figure}
    \centering
    \includegraphics[trim={32mm 15mm 40mm 26mm},clip, width=\textwidth]{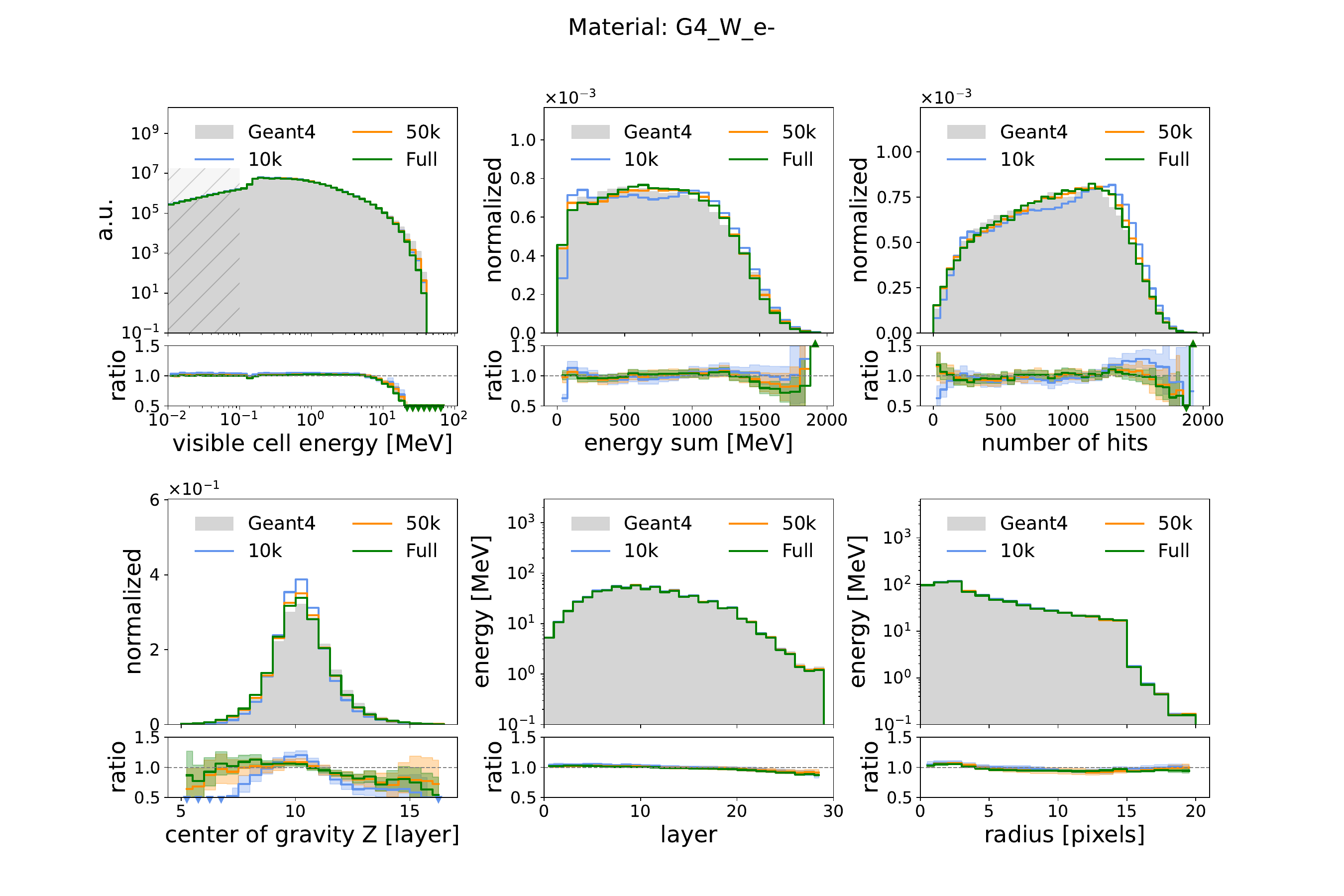}
    \caption{\textbf{Transfer learning efficiency for electron showers in tungsten across varying sample sizes:}
    Comparison between the ground truth \textsc{Geant4} reference (grey shaded) and the fine-tuned model at $10\text{k}$ (blue), $50\text{k}$ (orange), and full (green) sample sizes.
    Lower panels display the ratio to the \textsc{Geant4} reference, where a ratio of $1.0$ (dashed line) indicates perfect agreement. Shaded areas in the ratio plots depict the statistical uncertainty from data samples in quadrature with the uncertainty from bootstrapped fine-tunings, represented at the $3\sigma$ level.}
    \label{fig:e_tungsten}
\end{figure}

In this section, we evaluate the viability of fine-tuning—or more precisely, cross-species transfer learning—within the calorimeter. Adapting across particle types presents a greater challenge than transferring across detector materials, as the fundamental interaction dynamics can vary substantially. In our architecture, the attention layers encode interaction patterns established during pre-training, and since these are modified only via low-rank additive updates, the core representational structure of the base model remains relatively constrained. We are able to regain additional flexibility via particle-specific vocabulary heads; however, the constraining factors introduced by the narrow pre-training dataset remain.
The performance of this adaptation scheme is illustrated in Fig.~\ref{fig:e_tungsten}, following the same format as those in Fig. \ref{fig:gamma_Pb} in which we investigate the performance as a function of fine-tuning samples size.

Inspection of Fig. \ref{fig:e_tungsten} reveals excellent agreement between the Geant4 ground truth and the generated distributions. Specifically, the model demonstrates high-fidelity generation when adapted with 50k or more samples, closely matching the target physical distributions. While models trained on smaller datasets (e.g., 10k samples) remain consistent across most observables, they exhibit diminished accuracy in capturing specific shower characteristics, such as the center of gravity and total hit multiplicity.

\begin{figure}[ht]
    \centering
    \begin{subfigure}[b]{\textwidth}
        \centering
        \includegraphics[
            trim={32mm 15mm 40mm 26mm},
            clip,
            width=\textwidth,
            height=0.42\textheight,
            keepaspectratio
        ]{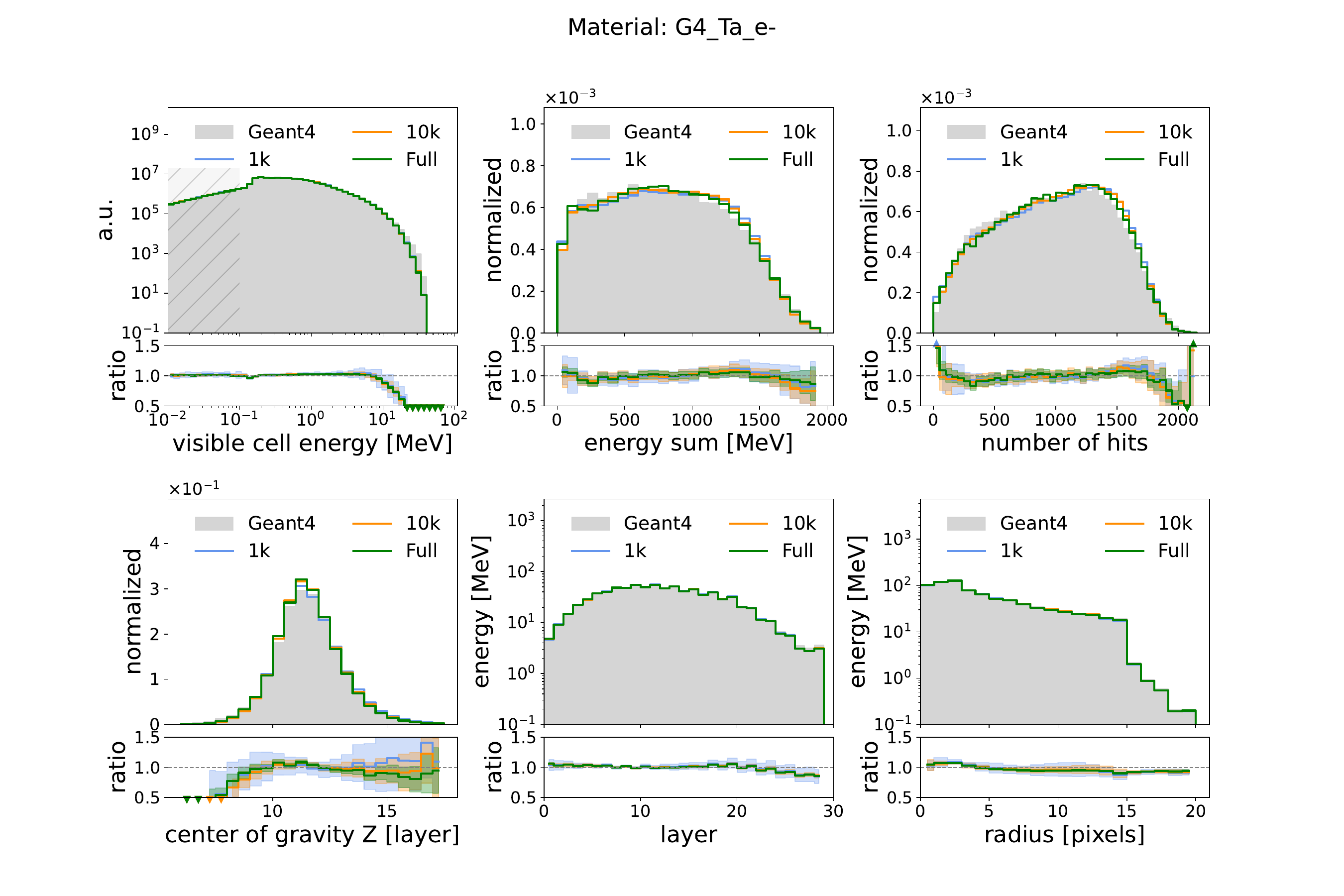}
        \caption{Tantalum}
        \label{fig:ta_electron}
    \end{subfigure}
    \vspace{0.01em}
    \begin{subfigure}[b]{\textwidth}
        \centering
        \includegraphics[
            trim={32mm 15mm 40mm 26mm},
            clip,
            width=\textwidth,
            height=0.42\textheight,
            keepaspectratio
        ]{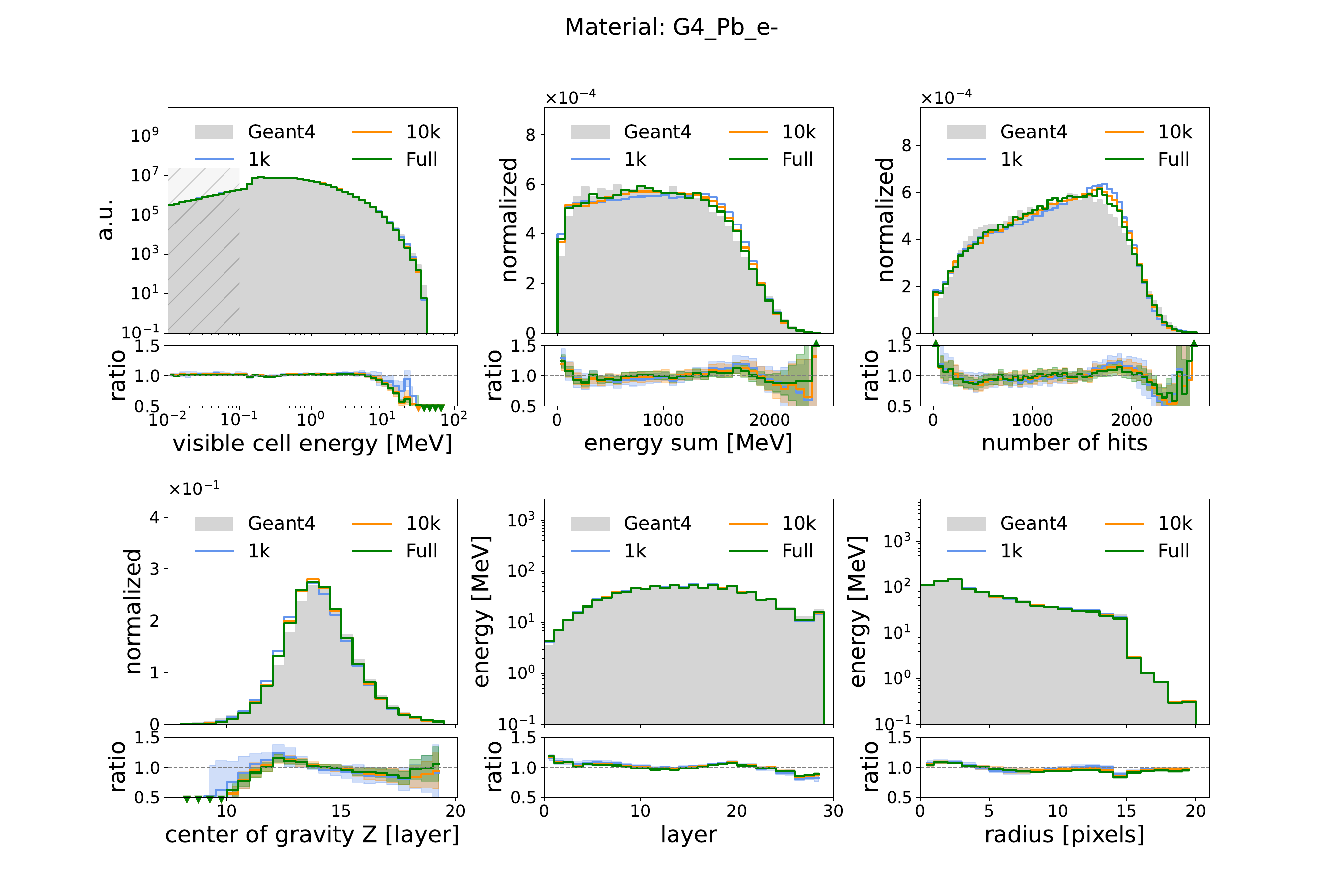}
        \caption{Lead}
        \label{fig:pb_electron}
    \end{subfigure}
    \caption{\textbf{Transfer learning efficiency for electron showers across varying sample sizes.}
    Comparison between the ground truth \textsc{Geant4} reference (grey shaded) and the fine-tuned model at $1\text{k}$ (blue), $10\text{k}$ (orange), and full (green) sample sizes for (a) tantalum and (b) lead.}
    \label{fig:electron_transfer_learning}
\end{figure}

We now validate the ability of our model to fine-tune to additional materials, \textit{e.g.}, electrons in tantalum and lead, given a prior adaption to a new particle species. To be specific, we now freeze the LoRA matrices and the electron specific vocabulary heads. Fine-tuning is performed adding and updating a single expert.
In this regard, the number of active parameters remains fixed, and therefore no additional inference cost is incurred compared to the electron-adapted model. The results are visualized in Figs. \ref{fig:ta_electron} and \ref{fig:pb_electron} for tantalum and lead, following an identical format as those previous.

Inspection of Figs.~\ref{fig:ta_electron} and \ref{fig:pb_electron} again shows excellent agreement between the ground truth and generated distributions. This agreement persists across all sample sizes, reflecting that tuning a single expert requires substantially fewer samples than the modular adaptations used previously for electron extension. Nevertheless, a clear systematic bias appears in the learned longitudinal shower profile. As the model is applied to materials further removed from the original transfer distribution (electrons in tungsten), the expert lacks sufficient capacity to perform the required shift in the shower development along the $z$ axis, and instead primarily captures the stochastic width of the shower profile for a given material. To correct for this effect, we apply a post-hoc adjustment in which the top-$k$ most energetic deposits are shifted deeper into the calorimeter, with $k=3$ for lead and $k=1$ for tantalum. This procedure can be interpreted as a calibration step applied to the model generations, and is consistent with prior generative approaches in which targeted post-processing---such as token duplication removal---is used to improve generation fidelity. We provide additional plots comparing distributions with, and without this calibration step in \ref{app:sys_bias}.

Given the fact that the transfer learning in this framework is performed through additive, modular modifications while the base model remains frozen, the adaptation capacity is fundamentally constrained compared to full fine-tuning, where the model parameters are free to move throughout the entire parameter space. When adapting the photon-trained model to electrons, the LoRA modules tend to organize themselves into a structured decomposition. The primary ranks act as an anchoring transformation that shifts the shower development forward in depth, reflecting the physical fact that electrons begin showering immediately, whereas photons typically require an initial conversion step. The remaining ranks then capture the stochastic components of the shower profile. This structure also propagates into the newly introduced vocabulary heads.

When this electron-adapted model is subsequently extended to new materials, the adaptation must contend with two competing constraints. First, the underlying base model still fundamentally encodes photon shower dynamics and has only been partially transformed to represent electrons through the limited expressive capacity of the LoRA modules. Second, these limited transformative features are themselves biased toward the initial material used during electron adaptation, as the low-rank structure can only encode a finite amount of information. As a result, extending the model to additional materials requires overcoming both the photon-dominated inductive bias of the base model and the material-specific bias introduced during the initial electron transfer step. This correction acts as a calibration of the generated showers and reflects a practical limitation of low-rank modular adaptation when further extrapolating across materials. We leave further study of methods for self correction within the model, or other external, model based corrective methods as future work.





\subsection*{Optimized Inference}

Our model achieves generation times comparable to normalizing flows \cite{papamakarios2021normalizingflowsprobabilisticmodeling} while maintaining the superior fidelity typically associated with transformer-based approaches \cite{CaloChallenge}.\footnote{We note that the scale of pre-training data used in this work is intentionally limited to match prior datasets and enable fair comparison. As a result, the full representational capacity of transformer-based models is not fully exploited in this setting.}
Table~\ref{tab:inference_speed} compares the inference speed of our model with several other models, using \textsc{Geant4} as the baseline.

\input{Tables/inference_speed_table}

Compared to \cite{birk2025omnijet}, the inference time of our model achieves a speedup of $\mathcal{O}(10^{3})$. Importantly, this improvement is not driven by architectural changes alone, but rather by the adoption of well-established inference optimization techniques from the LLM literature, including KV-caching, memory preallocation, and CUDA graph execution.

While the total parameter count of the model increases as the architecture expands its coverage (\textit{i.e.}, incorporating additional materials or particle species), the number of active parameters involved during generation remains fixed. In practice, only the shared backbone and the relevant expert and adaptation modules are invoked for a given configuration. As a result, the computational and memory requirements during inference remain approximately constant across different particle-material combinations, enabling scalable deployment without increasing per-instance inference cost.\footnote{
This is in good approximation given that LoRA introduces additional parameters, although these are
negligible in size compared to the base model.
}

Transformer-based autoregressive generation scales as $\mathcal{O}(n^2)$ in sequence length due to full self-attention recomputation at each token step. To address this, we implement a KV-cache \cite{pope2022efficientlyscalingtransformerinference} which stores the key and value projections from previously generated tokens, reducing per token step-complexity to $\mathcal{O}(n)$. This is particularly useful as the flattening of a normally 3-dimensional point cloud can lead to quite long 1-dimensional token sequences. 
The tradeoff to minimizing the self-attention computations is that the memory of the KV cache grows linearly with context length as $\mathcal{O}(n\times dim\times layers)$. We address this by preallocating the KV-cache to the maximum sequence length upfront, which trades the additional memory overhead for a fully static-allocation. This enables the usage of CUDA graph capture and eliminates per-step memory management costs during autoregressive generation. 

These techniques are well-established in LLM inference and are broadly applicable. Any autoregressive model with a transformer backbone, \textit{e.g.}, Omnijet-$\alpha_c$ \cite{birk2025omnijet}, could adopt KV-caching, preallocation, and CUDA Graphs to achieve comparable inference throughput.


%% file: Tables/inference_speed_table.tex
\begin{table}[ht]
    \centering
    \caption{\textbf{Inference Speed Comparison}: Comparison of inference speeds of various generative modeling approaches, and their relative speedup to \textsc{Geant4}. All generative modeling approaches are tested with identical hardware (Nvidia-A100). Inference time values for \textsc{Geant4}. Omnijet-$\alpha_c$, CaloCloudsII and L2LFlows are directly taken from \cite{birk2025omnijet}.}
    \label{tab:inference_speed}
    \resizebox{\textwidth}{!}{%
    \begin{tabular}{|l|c|c|c|c|}
        \hline
        \textbf{Model} & \textbf{Inference Time (ms)} & \textbf{Hardware} & \textbf{Speedup Factor} \\
        \hline
        \textsc{Geant4} \cite{GEANT4:2002, agostinelli2003geant4, AGOSTINELLI2003} & 4100 & CPU & N/A \\
        \hline
        OmniJet-$\alpha_c$ \cite{birk2025omnijet} & 3000 & A100 & 1.39 \\
        \hline
        CaloClouds II \cite{buhmann2024caloclouds} & 6.12 & A100 & 670 \\
        \hline
        L2LFlows \cite{L2LFlows} & 3.24 & A100 & 1260 \\
        \hline
        \textbf{Ours} & \textbf{10.46} & \textbf{A100} & \textbf{392} \\
        \hline
    \end{tabular}}
\end{table}

%% file: 5_summary.tex
\section{Summary and Conclusions}\label{sec:summary}

We present a foundation model for calorimetry capable of modeling multiple detector materials within a single architecture using a Mixture-of-Experts framework. 

The shared backbone enables efficient adaptation to additional materials through the introduction of lightweight expert modules, allowing new detector configurations to be incorporated with minimal modification to the base model. Importantly, this extension is performed in a modular manner that mitigates catastrophic forgetting and preserves alignment with the base model. Through a series of controlled studies, we demonstrate high-fidelity generation on both the base materials and those introduced through fine-tuning. Furthermore, we show that the model can be efficiently adapted in low-data regimes (\textit{e.g.}, 1k or 10k samples). 
%
This property is particularly valuable for detector design environments, as the foundation model can substantially reduce the number of new \textsc{Geant4} simulations required for modified detector configurations. Rather than generating large Monte Carlo samples from scratch, a pre-trained backbone can be efficiently re-adapted by training on a relatively small number of additional simulated events while preserving high-fidelity modeling of the detector response.
In addition, the architecture naturally supports conditional generation through the use of prepended contextual tokens, allowing next-token prediction to model particle-dependent shower development within a unified generative framework. In this work, we employ this mechanism to encode particle kinematics, although the same strategy can be extended to incorporate other design parameters that reside in continuous domains and cannot be naturally decomposed into discrete modular components.
The architecture further supports extension to new particle species through parameter-efficient fine-tuning. In particular, we employ additive and reversible low-rank adaptations that provide targeted corrections to the base representations while keeping the backbone frozen. The additive corrections are augmented through lightweight expert modules and particle-specific vocabulary heads, further enabling the model to disentangle particle-dependent shower characteristics from material-dependent effects. Once these particle-specific LoRA and vocabulary components are trained, further adaptation to additional detector materials follows the base model paradigm: the backbone, LoRA parameters, and particle vocabulary heads remain fixed, while only a new material expert is introduced and trained. 

This modular design enables efficient compositional adaptation across both particle species and detector materials without requiring retraining of the full model. This behavior is demonstrated through a series of closure tests in which we establish two key results. First, we show that initial adaptation to a new particle species (\textit{e.g.}, electrons) in tungsten achieves both high-fidelity transfer learning and strong sample efficiency, where only limited data (\textit{e.g.}, 50k) is required to successfully shift the model to the new particle domain. Second, we demonstrate that once these particle-specific modules have been trained, further adaptation to additional detector materials can proceed following the base-model fine-tuning paradigm, requiring only the introduction of a single lightweight expert module. This result directly supports integration into detector design pipelines, where multiple detector configurations and particle types must be evaluated to identify optimal design trade-offs.

Finally, we demonstrate that modern inference optimization techniques developed in the Large Language Model literature can be effectively applied to transformer-based simulation models in high-energy physics. In particular, by leveraging standard strategies such as key-value caching, memory preallocation, and CUDA graph execution, we significantly reduce the computational overhead associated with autoregressive generation. These optimizations eliminate redundant recomputation during decoding and improve GPU utilization, enabling efficient sequential inference even for long shower sequences. As a result, the effective inference throughput of our model becomes comparable to that of more traditional generative approaches used in fast detector simulation. This demonstrates that transformer-based generative models, when paired with established systems-level optimizations, can achieve practical inference performance and are therefore viable for deployment in physics simulation environments.

%% file: 6_Appendix.tex
\appendix

\section{Fixed Point Kinematic Generations} \label{app:fixed_point}

\begin{figure}[ht]
    \centering
    \includegraphics[trim={32mm 15mm 40mm 28mm},clip,width=\textwidth]{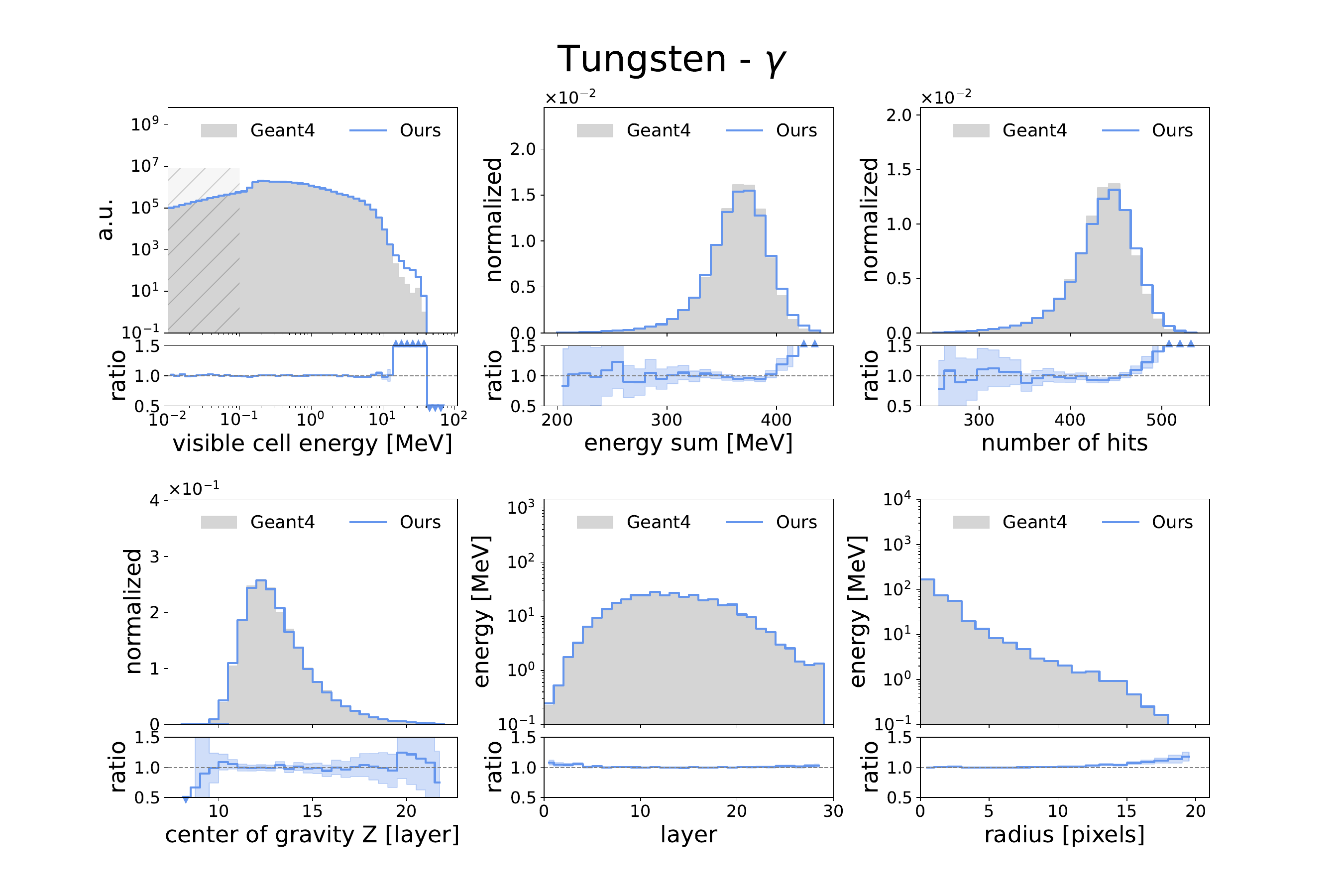}
    \caption{\textbf{Validation of energy kinematic conditioning through prepended context at 20 GeV}:
    Distribution level comparison between ground truth \textsc{Geant4} reference sample (gray shaded) and the proposed method (blue) at a fixed initial energy of $\SI[per-mode=symbol]{20}{\giga\eV}$ for photons in tungsten.
    }
    \label{fig:gamma_W_20GeV}
\end{figure}

\begin{figure}
    \centering
    \includegraphics[trim={32mm 15mm 40mm 28mm},clip,width=\textwidth]{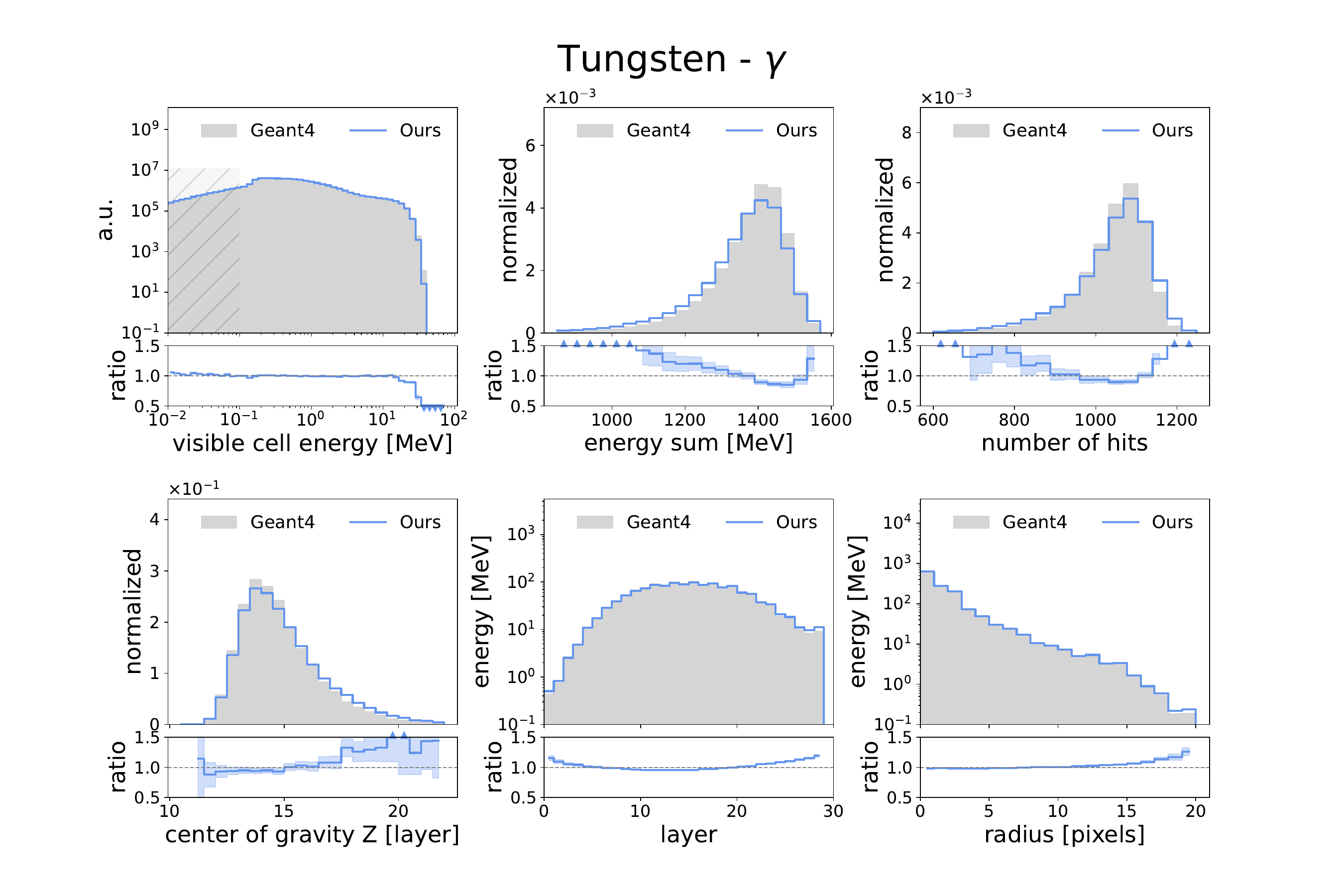}
    \caption{\textbf{Validation of energy kinematic conditioning through prepended context at 80 GeV}:
    Distribution level comparison between ground truth \textsc{Geant4} reference sample (gray shaded) and the proposed method (blue) at a fixed initial energy of $\SI[per-mode=symbol]{80}{\giga\eV}$ for photons in tungsten.
    }
    \label{fig:gamma_W_80GeV}
\end{figure}

We provide additional validation of our conditional generation scheme based on prepended contextual tokens. Figures~\ref{fig:gamma_W_20GeV} and \ref{fig:gamma_W_80GeV} present histogram-level evaluations for photon showers in tungsten at $\SI{20}{\giga\eV}$ and $\SI{80}{\giga\eV}$, respectively. These energies lie near the lower and upper bounds of the kinematic range considered in this study ($\SI{10}{\giga\eV}$--$\SI{100}{\giga\eV}$), thereby probing the model performance toward the extremes of the conditioning distribution.

\clearpage

\section{Bias Calibration}\label{app:sys_bias}

\begin{figure}[h]
    \centering
    \includegraphics[
        trim={32mm 15mm 40mm 26mm},
        clip,
        width=\textwidth,
        keepaspectratio
    ]{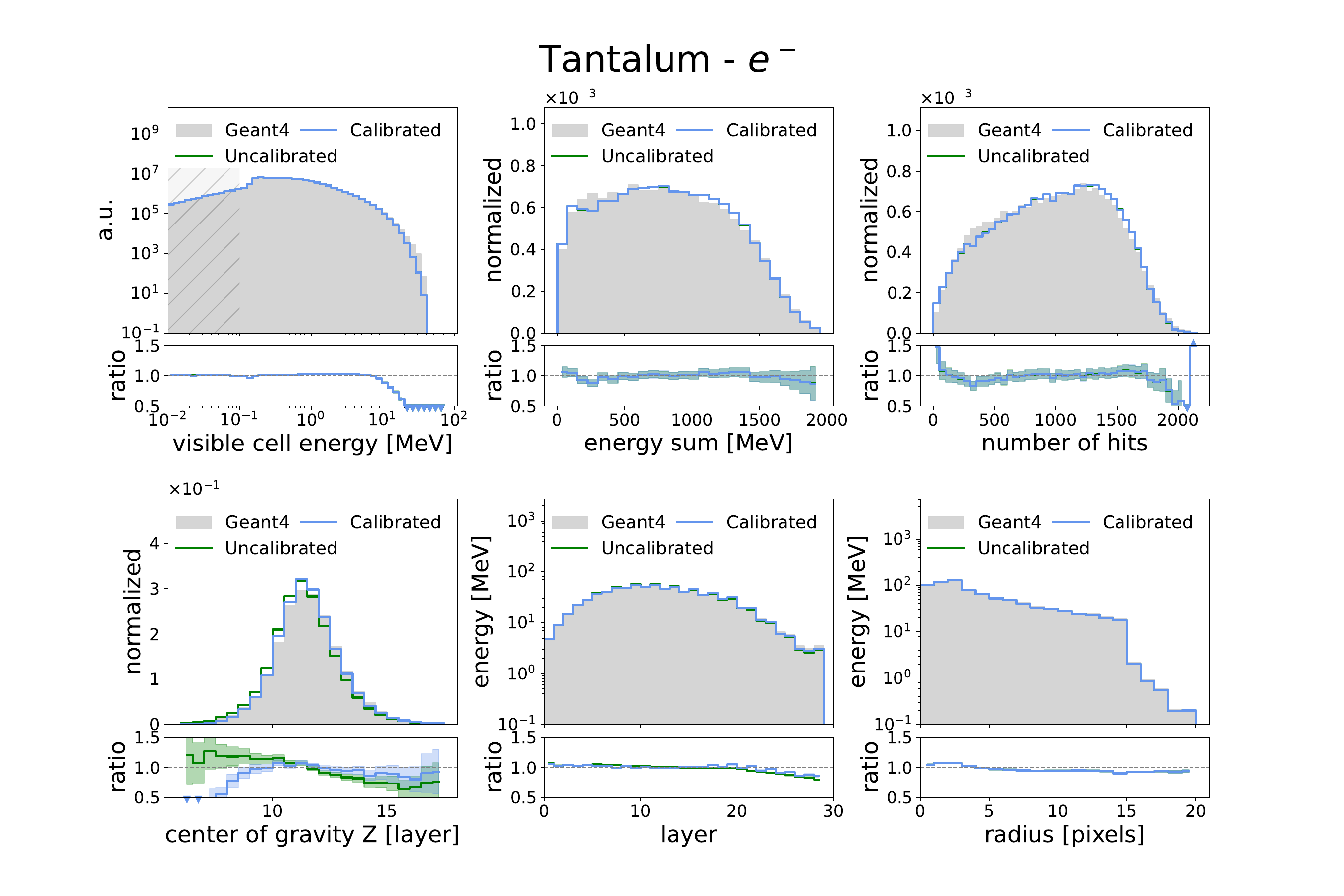}
    \caption{\textbf{Effect of calibration step on generated electrons in tantalum.}
    Comparison between the ground truth \textsc{Geant4} reference (grey shaded), the calibrated generations (blue), and the uncalibrated generations (green).}
    \label{fig:ta_electron_bias}
\end{figure}

\begin{figure}[h]
    \centering
    \includegraphics[
        trim={32mm 15mm 40mm 26mm},
        clip,
        width=\textwidth,
        keepaspectratio
    ]{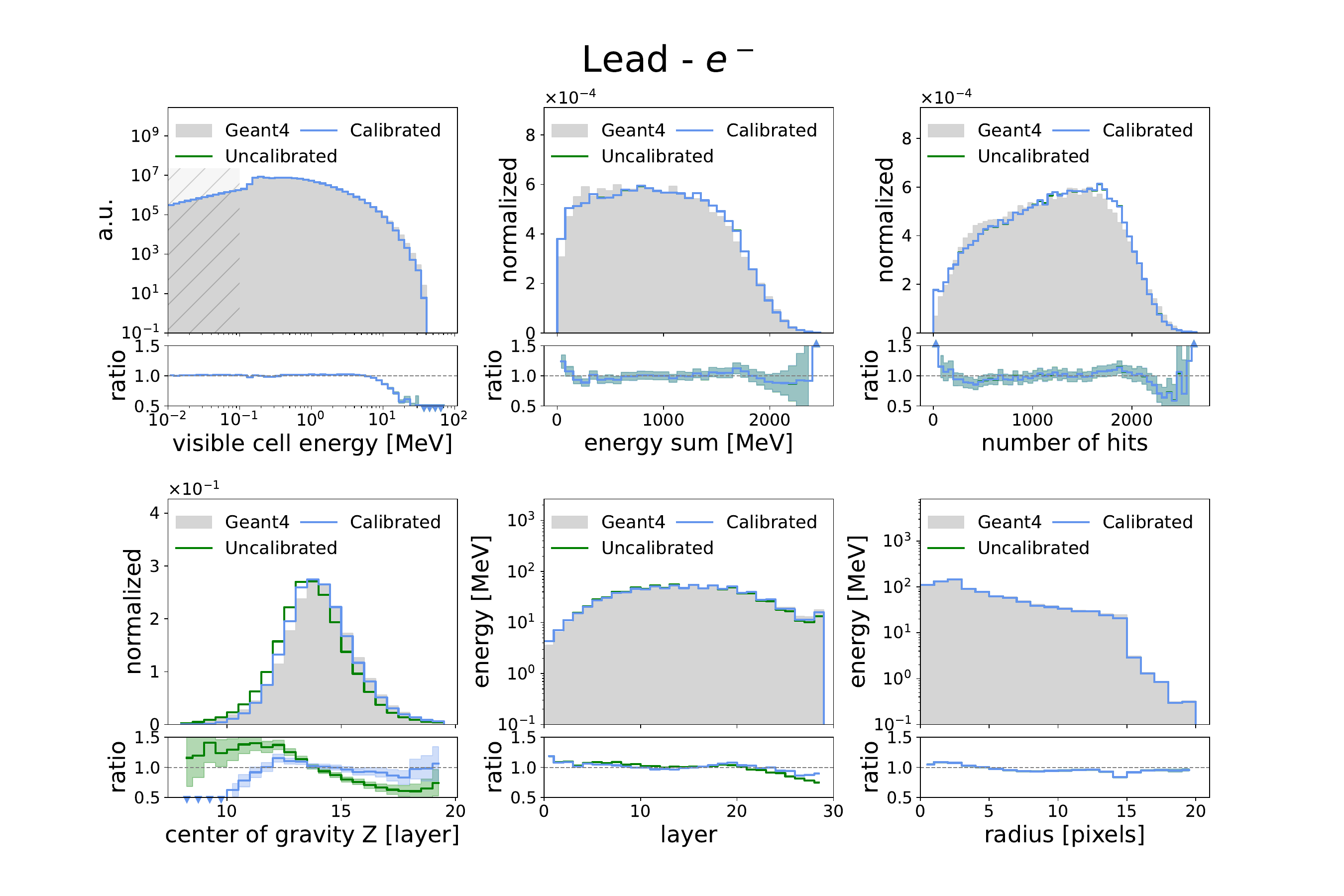}
    \caption{\textbf{Effect of calibration step on generated electrons in lead.}
    Comparison between the ground truth \textsc{Geant4} reference (grey shaded), the calibrated generations (blue), and the uncalibrated generations (green).}
    \label{fig:pb_electron_bias}
\end{figure}

We show additional comparisons illustrating the effect of the calibration step on the generated electron showers. Figures~\ref{fig:ta_electron_bias} and \ref{fig:pb_electron_bias} compare the calibrated and uncalibrated generations against the \textsc{Geant4} reference for tantalum and lead, respectively.

